# On the many '3-kiloparsec arms' – shocked wave and nuclear rotation

Jacques P. Vallée

National Research Council Canada, National Science Infrastructure, Herzberg Astronomy & Astrophysics, 5071 West Saanich road, Victoria, B.C., Canada V9E 2E7,
jacques.vallee@nrc-cnrc.gc.ca

**Abstract.**

Many features near the Galactic Center have been called "3-kiloparsec arms". We reached a point of having too many divergent data, making it difficult to be constrained by a single physical model.  Their differing characteristics suggest different physical and dynamical objects.

Radial velocity data on the so-called '3-kpc arms' do not coincide with radial velocities of major spiral arms near 3kpc, nor near 2 kpc, nor near 4 kpc from the Galactic Center (Fig. 1 and 2).

Different '3-kpc arm' features may require  different models: turbulence around a shock in a Galactic density wave between 2 and 4 kpc from  the Galactic Center (Table 1),  or nuclear rotation between 0 and 2 kpc from the Galactic Center region (Table 2), or a putative radial expansion between 0 and  4 kpc from the Galactic Center.

Despite their naming as 'Near 3-kiloparsec arms' or 'Far 3-kiloparsec arms', these features are not major arms. Those '3-kpc arms' features nearer the Galactic Center (within 13° of Galactic longitude) may be different than those farther out (Table 2).

Here we show that the plethora of observed '3-kpc arm' features can be separated in two: those with Galactic longitude of 13 degrees or more away from the Galactic Center (Table 1 - some of which are possibly associated with the observed major spiral arms), and those within 13 degrees from the Galactic Center (Table 2 - some of which are possibly associated with the observed central bars; Fig.1 and Fig.2).

## 1.  Introduction

The structure of the inner Galaxy was discussed in an earlier paper, covering the Galactic short boxy bar and bulge, the thin long bar, the spiral arm origins, and some '3-kpc arm' features (Vallée 2016a, and references therein).

Previous efforts were made to identify and physically understand the '3-kpc arm' features. The identification of the '3-kpc arm' features dates back to van Woerden et al (1957) and Oort et al (1958). Over the next 60 years, there were many following research works, notably by Bania (1980), Hayakawa et al (1981), Beuermann et al (1985), Sevenster (1999), Dame & Thaddeus (2008), Churchwell et al (2009), Green et al (2009; 2011), Caswell et al (2010), Sanna et al (2009, 2014).

The nature or physical understanding of the '3-kpc arm' features has not converged yet. It could be an expanding ring structure (e.g., van der Kruit 1971; Chen & Davies 1976, Sevenster 1999), flowing elliptical streamlines (e.g., Peters 1975), or expanding spiral arms (Fux 1999, Bissantz et al 2003).

The number of papers on the subject of the '3-kpc arms' keeps increasing. The '3-kpc arms' are observed in gas motion in velocity space. Yet fundamental parameters remain unanswered, such as a proper tangent from the Sun to some of these objects in Galactic longitude, or a proper radial distance from the Sun.

Here we employ recent studies on the origins of spiral arms, in order to look for answers on the tangent longitude, radial distance, and proper origin of these '3-kpc arms'. In Section 2 we look for possible velocity location or separation of spiral arms and of the '3-kpc arms'. In Section 3 we look for tangents in Galactic longitudes to these '3-kpc arms', and a possible Galactic model. In Section 4, we prove that these '3-kpc arms' cannot be arms. In Section 5 we look for two other possible but different models for the '3-kpc arms'.

The motivation for this work stems from the observations of discrepant data that have accumulated on the so-called '3-kpc arms', being incompatible with a single physical model, and attempts to **separate** these different data into different radial velocities (Figures 1 and 2), different Galactic longitudes (Table 1) and different attributes (Table 2), and to link them potentially with different physical models (turbulence in a shocked wave, or nuclear rotation).

## 2. New numerical processing

In earlier papers (Vallée 2016a; Vallée 2016b), we extended our spiral arm model to go deeper toward the Galactic Center [GC], based on very recent observational data in Galactic longitudes.

Here we improved on our earlier velocity model (Vallée 2008) by fine-tuning its parameters, namely its pitch angle (-13.1$^o$ ; Vallée 2015), the start of each spiral arms (2.2 kpc; Vallée 2016a), the Galactic longitude tangent to each spiral arm (Tables 3 to 10 in Vallée 2016b). In addition, we used the most recent value for the orbital velocity of the local standard of rest going around the Galactic Center – two dozens measurements of $V_{lsr}$, as published from mid-2012 to early-2017 indicated a median and mean for $V_{lsr}$ close to 230 ± 3 km/s (Vallée, 2017). Also, we used the most recent value for the distance of the Sun to the Galactic Center - the review by De Grijs & Bono (2016) yielded $R_{sun}$ = 8.3 ±0.4 kpc, that of Gillessen et al (2013) showed a median near 8.1 ±0.3 kpc, that of Malkin (2013) found a weighted mean value of 8.0 ± 0.4 kpc, Vallée (2017) found a weighted mean of 8.0 ± 0.2 kpc, and one gets an unweighted arithmetic mean of 8.0 ±0.4 kpc in Table 3 of Bland-Hawthorn & Gerhard (2016).

Numerical results will appear elsewhere for the whole Galaxy; here we concentrate on the inner Galaxy, pertaining to the area encompassing the so-called '3-kpc arms'. The model used is that of Vallée 2016a – Fig.2), but with two different circular velocities (see below).

The spiral model employs the start of each spiral arm at 2.2 kpc from the Galactic Center, as needed to match the recently observed spiral arm tangents for the inner spiral arms (Vallée 2016a – fig.4; Vallée 2016b – table 3) and a short boxy bar (Vallée 2016a – table 3).

Some earlier models employed a farther start for the arm origin (3kpc or 4 kpc), a different arm pitch angle, and a long bar (4 kpc radius) to then predict the inner Galaxy (Zhang et al 2014 – fig.8). Yet the importance of the long bar has been questioned, as it crosses the inner spiral arms without perturbing them (section 3.2 in Vallée 2016a). Observations of the inner Galaxy are still in progress, so the adopted spiral arm model in this paper is reasonable.

**Figure 1** shows a zooming of the arms in Galactic quadrant IV, within 30º of the Galactic Center longitude (l=0º).

The filamentary features known as '3-kpc arms' are drawn from longitude -12º up to longitude +12º, as observed in Fig. 1 in Dame & Thaddeus (2008); the near one is from (l,v)= 348º, -103 km/s up to (l,v)= 13º, +0 km/s, and the far one is from (l,v)= 348º, +8 km/s up to (l,v)= 13º, +108 km/s.

**Figure 2** shows a zooming of the arms in Galactic quadrant I, within 30º of the Galactic Center longitude (l=0º).

The mean orbital velocity of 230 km/s found earlier (see Section 2) has a standard deviation of the mean of 3 km/s, but a root-mean-square deviation of 11 km/s from the mean. Fig.1a and Fig.2a show the orbital velocity of 220 km/s, while fig.1b and 2b show the orbital velocity of 240 km/s. The readers will chose a curve in the statistically permitted range shown.

As shown in Figures 1 and 2, the spiral arms and these '3-kpc arms' appear to be independent in velocity – longitude plots (near longitude $0^o$). The arms do not seem to have an obvious overlap or a gross interference with these objects, within $30^o$ of the Galactic Center.

We confirm here the absence of dependence in velocity space between these '3-kpc arms' and the regular spiral arms in the inner Galaxy. This confirmation follows from using the start of the spiral arms at around 2.2 kpc from the Galactic Center, close to where the '3-kpc arm' features are observationally found (Vallée 2016a). No such velocity analysis could be made if other models are employed with the start of the spiral arms farther away, between 3 and 4 kpc from the Galactic Center (outside the radius range of the '3-kpc-arm' features).
The spiral arm used (Fig. 2 in Vallée 2016a) are converted into a longitude – velocity diagram (Fig. 1 and 2 here) by the use of a rotation curve: a flat one with a circular velocity at the sun of 220 km/s (fig. 1a and 2a), and another flat one with the circular velocity being 240 km/s (fig.1b and 2b). As mentioned earlier, there is a root-mean-square deviation of 11 km/s from the mean of 230, so the readers can choose a curve in the statistically permitted range shown (220 to 240).

The '3-kpc-arm' features shown longitude-velocity in Fig. 1 and 2 are shorter in longitude and space than that of long spiral arms, and having *different* velocities than spiral arms (hence a different dynamical origin). Farther away from the Galactic Center, observations of our Galaxy and of other galaxies show some short arm spurs or arm branches (armlets) located in between long spiral arms, but with *similar* velocities as their nearby spiral arms (hence armlets are dynamically different than '3-kpc arm' features)..

### 3. Tangents to the '3-kpc arms'

The '3-kpc arms' are seen within $27^o$ of the Galactic Center, yet finding a tangent from the Sun to these objects is problematic.

**3.1** No tangent predictions within $\pm 13^o$ of the Galactic Center

A literature search for the line-of-sight tangents from the Sun to the so-called '3-kpc arms' reveals that there are more predictions than observational data.

Using HI at 21cm wavelength, Van Woerden et al (1957) did not find an arm tangent to the high-velocity filament, when mapping from new Galactic longitude l= $-10^o$ to l= $+3^o$ (their fig.3), adding: "ce bras n'est pas vu tangentiellement" (this arm is not seen tangentially).

Using CO 1-0 data, Bania (1980) observed the 3-kpc arm feature as a line extending from (l,v)= 350°, -100 km/s up to (l,v)= 14°, +25 km/s – their Section IIIb and their Table 4, and concluded that the '3-kpc arm' feature is not a continuous structure; no tangent to the feature is observed.

Using CO 1-0 data, Dame and Thaddeus (2008) predicted the tangents for the '3-kpc arms' to be at l= -23° and l= +23° (corresponding to about 3 kpc from the Galactic Center), although their CO observational data cover only from -12° < l < +13° (about 1.8 kpc from the Galactic Center).

Thus, there were no tangents to these '3-kpc arms' observed between Galactic longitudes -13° to +13°.

**3.2** Many tangents away from the GC, with links to spiral arms

It was shown elsewhere the tangents to these '3-kpc arm' filamentary features could be very close to the tangents of known spiral arms (table 5 and figure 5 in Vallée 2016a), namely, close to the inner Perseus arm (l ≈ -23°), close to the inner Sagittarius arm (l≈ -17°), close to the inner Norma arm (l≈ +20°), or close to the inner Scutum arm (l≈ +30°). This is supported globally by the reversal in Galactic longitudes, across the Galactic Meridian, of the chemical tracers of each arm (Fig. 1 in Vallée 2016b; Vallée, in preparation).

The start of each spiral arm near the Galactic nucleus is not yet final, as observations are difficult at that distance. Extensions of the arms from the solar distance to deep in the inner Galaxy depend on good fits to observational data outside the inner Galaxy, notably on fitting the longitude of each arm tangent as seen from the Sun.

The choice of the start of the inner spiral arm (2.2 or 4 kpc), and the choice of the short versus the long Galactic bar (2.1 or 4.2 kpc), must be made together along with the observed relative Galactic longitude positioning of the chemical different tracers for the inner Norma spiral arm, as discussed earlier in Vallée (2016a- Section 5 and Table 3; Vallée 2016b -Table 3); this explains the presence of the dichotomy presented here in Table 2.

Some other models do not extends arms that close to the Galactic Center, as they prefer to extend the model Galactic bar to 4 kpc from the Galactic Center (e.g., fig. 8 in Zhang et al 2014; Fig. 1 in Reid et al 2014) and attach the Scutum arm to this long bar near 4 kpc near l=+32° (yet the Scutum arm shows up observationally near near l=+26° for dust to +32° for CO, see Table 3 in Vallée 2016b); it follows that the long bar would eliminate the inner Norma spiral arm (no inner Norma arm is physically possible, since it would cross and disrupt the long bar near l=+16° for masers and l=+20° for CO). This conundrum was discussed in Vallée (2016a – section 3).

**Table 1** accumulates published data on the longitudes of these features, possibly associated with spiral arms. There is a void of predictions of the tangent longitudes between -13° and +13° in Table 1.

### 3.3 Summary and possible interpretation

If these '3-kpc arms' located away from the Galactic Center (Table 1) are associated or linked to spiral arms, they may have an origin that is associated with turbulence around a shocked spiral arm, such as a shock-induced wiggle instability and swing amplification around an arm formed by the Galactic density wave (Shu 2016 – Section 3.2 and Section 5.7).

A definite measurement of the radial distance of these '3-kpc arms' could help differentiate between attachment model to the Perseus and Scutum spiral arms (3 to 4 kpc) and the attachment model to the inner start of the Sagittarius and Norma spiral arms (2 to 3 kpc).

**Table 2** summarizes the attributes of the '3-kpc arms' and the spiral arms, and splits them in two pairs (one close to, and one far from, the Galactic Center). There are many divergences between the two objects, yet too few similarities.

### 4. Impossible spiral arm model for the '3-kpc arms'

Here we compare the '3-kpc arms' with the standard spiral arm model (see Fig.2 in Vallée 2016a).

The '3-kpc arms' are not similar to a known major spiral arm. Here is what the '3-kpc arms' are not: not spiral-shaped, not a continuous feature, not at a spiral arm's velocity range, no line-of-sight tangent close to the GC, not observed at the Galactic distance greater than 4 kpc; no OH maser (6.0 GHz) is seen at all along the '3-kpc-arms' (grey lines in Fig. 4 of Avison et al 2016) while these masers are seen in all spiral arms;

Also there are no large 'wiggles' nor curvature in Galactic longitude seen due to the Galactic bar (none seen in Figure 1 in Dame & Thaddeus 2008). The boxy bulge bar extending to a radius of 2.1 kpc at a position angle of 23° (Table 3 in Vallée 2016a) does not seem to affect the '3-kpc arms' in CO between -12°< l < 12° (Fig. 1 in Dame & Thaddeus 2008; Fig.7 in Sanna et al 2014). There is little or no wiggle in a range of Galactic longitude, and little or no wiggle in LSR in a range of radial velocity.

Hence for these many reasons, we confirm that the '3-kpc arms' are *not* following a proper model of major spiral arms.

Could the '3-kpc arm' features be created in the same way as short arm spurs, or branches, or armlets, seen in some other spiral galaxies? As mentioned above, the '3-kpc arm' features have *different* velocities than adjacent spiral arms while short arm spurs or arm branches or armlets located in between long spiral arms have *similar* velocities as their adjacent spiral arms. Hence it is possible to think of their dynamical origins as sufficiently different.

### 5. Models for the '3-kpc arms' near the Galactic Center

Excluding the tangents to the spiral arms, the other observables within $13^o$ of the Galactic Center require another physical model (not a Galactic density wave).

**5.1** The '3-kpc arms' and the expansion model up to 3kpc

The different velocity of the '3-kpc arms' is puzzling. The absence of a proper arm model, and the absence of large 'wiggles' or curvature has favored 'expansion' models, predicted to be near a Galactic radius of 3 kpc, perhaps smaller (van Woerden et al, 1957; Dame & Thaddeus 2008). An expanding ring is often cited (Sevenster 1999).

This expansion model agrees with some apparently large Galactic longitude offsets from the Galactic Center. The large offsets in Galactic longitudes may need an expansion model or else the turbulence model around spiral arms as predicted by the shocked Galactic density wave theory could happen.

**5.2** The '3-kpc arms' and the nuclear rotation model up to 2 kpc

A rotating bar model, suitably adapted, could perhaps explain the near '3-kpc arm' feature from (l,v)= $348^o$, -103 km/s up to (l,v)= $13^o$, +0 km/s, and the far '3-kpc arm' filamentary feature from (l,v)= $348^o$, +8 km/s up to (l,v)= $13^o$, +108 km/s (they appear to fit very nicely inside the green orbits in Fig. 8 of Habing 2016).

Within $12^o$ of the longitude $0^o$, there are several maser stars that are found at high radial velocities above 220 km/s (Fig. 2 in Habing 2016); the gravitational potential of a rotating Galactic bar was proposed to adequately explain their longitude and velocity.

If we follow a rotating-bar model for the '3-kpc arms', then these features are all located within 2.2 kpc of the Galactic Center (not at 3 or 4 kpc away). This rotation model could perhaps fit an inner pair of '3-kpc arms', provided their distances are nearer to 2 kpc.

### 6. Conclusion

Many of the '3-kpc arms' have unknown origins, distances, and Galactic tangents. Our results here indicated that they are definitely not following a velocity or longitude model of a proper spiral 'arm', despite each one being misnamed as a

'3-kpc arm'. In the inner Galaxy, the '3-kpc arms' are examined for a possible relation to the inner spiral arms (Fig. 1 and 2). None are found in velocity space – see Section 2. This area covers the radius range of both the '3-kpc-arm' features and of the start of the spiral arms at around 2.2 kpc from the Galactic Center.

Some of these features seen at large longitudes from the Galactic Center might appear to be loosely connected to the starts of the arms (at 2 kpc) or later along the arms (at 4 kpc), as explained in Table 1. Turbulences generated by the Galactic density waves could explain some '3-kpc arms' features near major spiral arms (Shu et al 2016, sections 3.2 and 5.7).

There seem to be two different sets of '3-kpc arms', each at different Galactic longitudes (one at a small offset and one at a large offset from the Galactic Center – see Table 2).

The '3-kpc arms' are misnamed, as they are not 'arms' (see Section 4). A definite measurement of the radial distance of these '3-kpc arms' could help differentiate between the expansion model or the turbulent density wave model (for '3-kpc arms' located from 2 to 3 kpc of the GC), and the expansion model or the nuclear rotation model (for '3-kpc arms' located from 0 to 2 kpc of the GC). At the limit, one could envision a time when there would be no need for the radial expansion model, if all data can be covered by the other models (Table 2).

**Acknowledgements.**
The figure production made use of the PGPLOT software at NRC Canada in Victoria. I thank an anonymous referee for useful, careful, and historical suggestions.

**References**
Avison, A., et al.: 2016, MNRAS, 461, 136.
Bania, T.M.: 1980, ApJ, 242, 95.
Beuermann, K., Kanbach, G., Berkhuijsen, E.: 1985, A&A, 153, 17.
Bissantz, N., Englmaier, P., Gerhard, O.: 2003, MNRAS, 340, 949.
Bland-Hawthorn, J., Gerhard, O.: 2016, ARAA, 54, 529.
Caswell, J.L., Fuller, G.A., Green, J.A., et al: 2010, MNRAS, 404, 1029.
Churchwell, E., Babler, B.L., Meade, et al: 2009, PASP, 121, 213.
Cohen, R.J., Davies, R.D.: 1976, MNRAS, 175, 1.
Dame, T., Thaddeus, P.: 2008, ApJ, 683, L143.
De Grijs, R., Bono, G., 2016, ApJ.Sup.Ser., 227, 5.
Fux, R.: 1999, A&A, 345, 787.
Gillessen, S., et al: 2013, Proc. IAU Symp., 289, 29.
Green, J.A., McClure-Griffiths, N.M., Caswell, J.L., et al: 2009, ApJ, 696, L156.
Green, J.A., Caswell, J.L., McClure-Griffiths, N.M., et al: 2100, ApJ, 733, 27.
Habing, H.J.: 2016, A&A, 587, 140.
Hayakawa, S., et al: 1981, A&A, 100, 116.

Malkin, Z.M., 2013, Astron. Rep., 57, 128.
Oort, J.H., Kerr, F.J., Westerhout, G.: 1958, MNRAS, 118, 379.
Peters, W.L.: 1975, ApJ, 195, 617.
Reid, M.J., Menten, K.M., et al: 2014, ApJ, 783, 130.
Sanna, A., Reid, M.J., et al: 2009, ApJ, 706, 464.
Sanna, A., Reid, M.J., et al: 2014, ApJ, 781, 108.
Sevenster, M.N.: 1999, MNRAS, 310, 629.
Shu, F.H.: 2016, Annual Rev. Astron. & Astrophys., 54, 667.
Vallée, J.P.: 2008, AJ, 135, 1301 .
Vallée, J.P.: 2015, MNRAS, 450, 4277.
Vallée, J.P.: 2016a, Astron J, 151, 55.
Vallée, J.P.: 2016b, ApJ, 821, 53.
Vallée, J.P.: 2017, Astrophys. Space Sci., 362, 79.
Van der Kruit, P.C.: 1971, A&A, 13, 405.
van Woerden, H., Rougoor, G.W., Oort, J.H.: 1957, Comptes Rendus Acad Sci. Paris, 244, 1691.
Zhang, B., Moscadelli, L., et al: 2014, ApJ, 781, 89.

**Table 1 – Some '3-kpc arms' in various Galactic Quadrants (GQ), longitudes and radial distances from Galactic Center (GC).**

| GQ no. | Gal. longitudes degree | GC dist. kpc | Associated arm at similar Gal. longitude |
|---|---|---|---|
| IV | -22 to -24 | 3.0 to 3.3 | tangent to 'inner Perseus arm' near -23° |
| IV | -13 to -21 | 1.5 to 2.9 | tangent to 'inner Sagittarius arm' at -17° |
| I  | +13 to +24 | 1.7 to 3.3 | tangent to 'inner Norma arm' near +20° |
| I  | +25 to +27 | 3.4 to 3.6 | tangent to Scutum arm near +30° |

Note: Longitude references in Vallée (2016a – table 5); Galactic Center distances of '3-kpc-arm' features are read from figure 5 in Vallée (2016a), using 8.0 kpc as the Sun to GC distance.

**Table 2. Comparison of attributes to the arms and to the '3-kpc arms' (within or outside Galactic longitude |l| =13°)**

| Attribute | spiral arm | '3-kpc arms' within 13° of GC | '3-kpc arms' outside 13° of GC |
| --- | --- | --- | --- |
| Velocity | small to medium | high | high |
| Shape | spiral | discontinuous | discontinuous |
| Extent | long | short | short |
| Longitude | 360° | < 13° from GC | >13° from GC |
| Link to spiral arm | - | no | yes |
| Line of sight tangent | many | none | 2 to 4 kpc from GC |

**Figure captions**

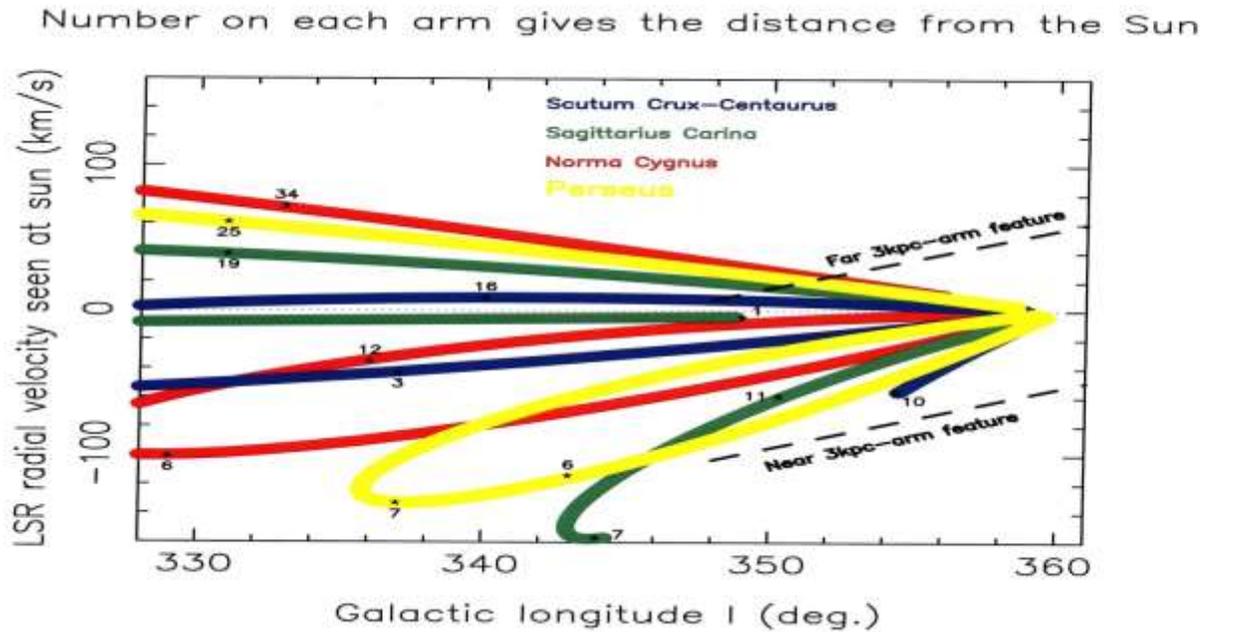

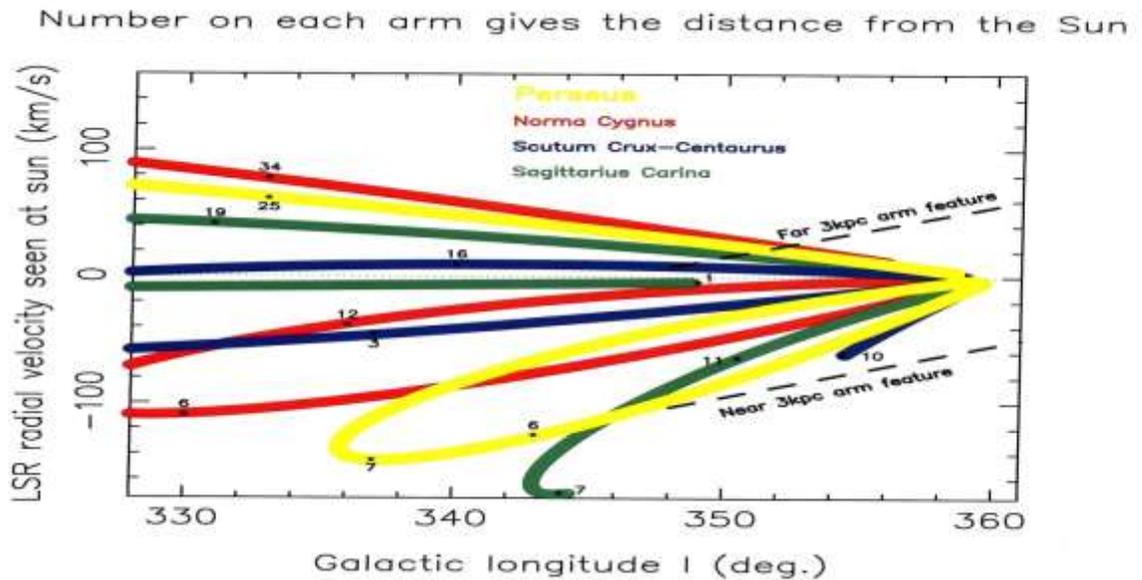

**Figure 1.** Zoom of the velocimetric model near the Galactic Center, in Galactic quadrant IV. Dashed lines refer to the filamentary feature called 'Near 3kpc arm' at negative velocities, and to the filamentary feature called 'Far 3kpc arm' at positive velocities. The number on each arm gives its distance from the Sun, assuming a Sun to Galactic Center distance of 8.0 kpc.
    **a)** l-v with $V_{lsr}$ = 220 km/s.      **b)** l-v with $V_{lsr}$=240 km/s.

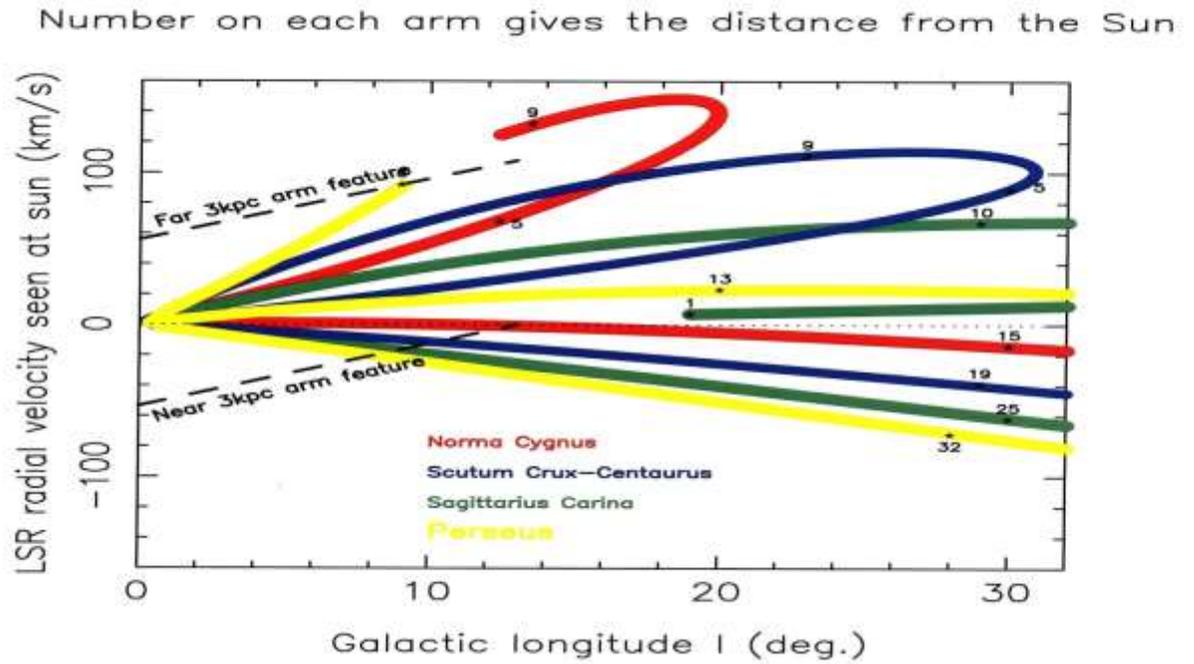

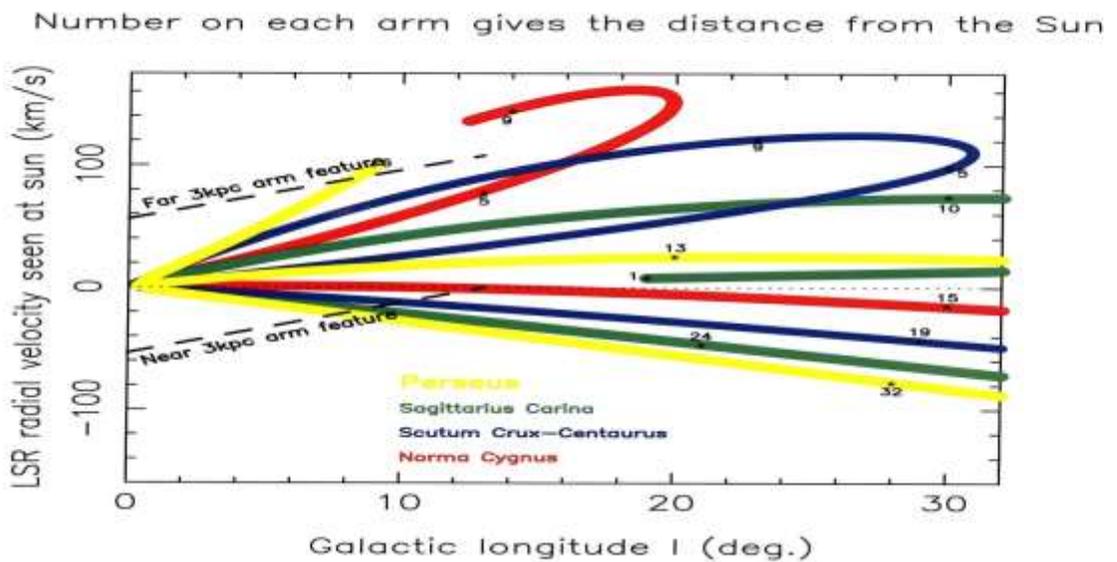

**Figure 2.** Zoom of the velocimetric model near the Galactic Center, in Galactic quadrant I. Dashed lines refer to the filamentary feature called 'Near 3kpc arm' at negative velocities, and to the filamentary feature called 'Far 3kpc arm' at positive velocities. The number on each arm gives its distance from the Sun, assuming a Sun to Galactic Center distance of 8.0 kpc.
**a)** l-v with $V_{lsr} = 220$ km/s.         **b)** l-v with $V_{lsr}=240$ km/s.